# Data Mining : A prediction of performer or underperformer using classification

Umesh Kumar Pandey    S. Pal

**VBS Purvanchal University, Jaunpur**

**Abstract** — Now a day's students have a large set of data having precious information hidden. Data mining technique can help to find this hidden information. In this paper, data mining techniques name Byes classification method is used on these data to help an institution. Institutions can find those students who are consistently perform well. This study will help to institution reduce the drop put ratio to a significant level and improve the performance level of the institution.

## I. INTRODUCTION

As the cost of processing power and storage is coming down, data storage became easier and cheaper so the amount of data stored in educational databases is increasing rapidly. In order to get benefits from such large database to find hidden relationships between variables using different data mining techniques developed and used.

Data mining, sometimes also called Knowledge Discovery in databases (KDD), can be very useful in the student-centered educational system. Within the KDD process, there can be used different means of data mining analysis that allow getting important information from the database such as: classification, clustering, association, decision tree, neural network etc [5].

Classification is perhaps the most familiar and most popular data mining technique. Predication can be thought of as classifying an attribute value into one of a set of possible classes. It is often viewed as forecasting a continuous value, wile classification a discrete value [10].

This study aims to analyze previous year data and predict New Year student as a performer and underperformer. These applications can help both instructor and student to enhance the quality education. The classification carried out using a Bayesian classification method.

## II. DATA MINING

Data mining techniques are used to operate on large volumes of data to discover hidden patterns and relationships helpful in decision making [3]. Alternatively it has been called exploratory analysis, data driven discovery and deductive learning. Data mining access of a database differs from this traditional access in several ways: query, data and output [10]. A data mining algorithm is a well-defined procedure that takes data as input and produces output in the form of models or patterns. The term well-defined indicate that the procedure can be precisely encoded as a finite set of rules [4]. The structures discovered during the data mining process can describe the entire (the most of the) set of data and they are called "models". There are also cases when the structures discovered get some local properties of the data and in that case the term of "pattern" is used [5]

## III. BACKGROUND AND RELATED WORK

Data mining is an emerging methodology used in educational field to enhance our understanding of learning process to focus on identifying, extracting and evaluating variables related to the learning process of students [2]. Data mining can be applied to a number of different applications such as data summarization, learning classification rules, finding associations, analyzing changes and detecting anomalies [8]. Educational data mining is used to identify and enhance educational process which can improve their decision making process [6]. Gabrilson uses the data mining prediction technique to identify the most effective factor to determine a student's test score, and then adjusting these factors to improve the student's test score performance in the following year [7]. Luan uses data mining to group students to determine which student can easily pile up their courses and which take courses for longer period of time [9].

## IV. CLASSIFICATION

Predictive modeling is the process by which a model is created or chosen to try to best predict the probability of an outcome. In many cases the model is chosen on the basis of detection theory to try to guess the probability of an outcome given a set amount of input data [16]

Classification is a predictive data mining technique, makes predication about values of data using know results found from different data [10. Predictive models have the specific aim of allowing us to predict the unknown value of a variable of interest given known values of other variables. Predictive modeling can be thought of as learning a mapping from an input set of vector measurements x to a scalar output y [4]. Classification maps data into predefined groups are classes. It is often referred to as supervised learning because the classes are determined before examining the data. They often describe these classes by looking at the characteristic of data already known to belong to the classes [10].

## V. BAYESIAN CLASSIFICATION

Bayes classification has been proposed that is based on Bayes rule of conditional probability. Bayes rule is a technique to estimate the likelihood of a property given the set of data as evidence or input Bayes rule or Bayes theorem is-

$$P (h_i \mid x_i) = \frac{P ( x_i \mid h_i ) \ P( h_i )}{P (x_i \mid h_i) + P (x_i \mid h_2) \ P(h_2)}$$

The approach is called "naïve" because it assumes the independence between the various attribute values. Naïve Bayes classification can be viewed as both a descriptive and a predictive type of algorithm. The probabilities are descriptive and are then used to predict the class membership for a target tuple. The naïve Bayes approach has several advantages: it is easy to use; unlike other classification approaches only one scan of the training data is required; easily handle mining value by simply omitting that probability [10]. An advantage of the naive Bayes classifier is that it requires a small amount of training data to estimate the parameters (means and variances of the variables) necessary for classification. Because independent variables are assumed, only the variances of the variables for each class need to be determined and not the entire covariance matrix. In spite of their naive design and apparently over-simplified assumptions, naive Bayes classifiers have worked quite well in many complex real-world situations [16].

**Table 1**

| | | Division | | | |
|---|---|---|---|---|---|
| | | I | II | III | FAIL |
| **Caste Category** | GEN | 100 | 120 | 34 | 46 |
| | OBC | 61 | 78 | 19 | 12 |
| | SC/ST | 29 | 50 | 28 | 23 |
| **Language Medium** | ENGLISH | 100 | 70 | 18 | 6 |
| | HINDI | 90 | 178 | 63 | 65 |
| **Class** | BA(NC) | 12 | 76 | 54 | 22 |
| | BA(CA) | 43 | 72 | 4 | 10 |
| | BSc(Bio.) | 80 | 50 | 15 | 18 |
| | BSc(Math) | 54 | 20 | 6 | 8 |
| | Bcom | 1 | 30 | 12 | 13 |

## VI. EDUCATIONAL DATA MINING IN HIGHER EDUCATION

Providing higher education to all sector's of a nation's population means confronting social inequalities deeply rooted in history, culture and economic structure that influence an individual's ability to compete. Quality assurance in higher education has raised to the top of the policy agenda in many nations [12].

Educational Data Mining is an emerging discipline, concerned with developing methods for exploring the unique types of data that come from educational settings, and using those methods to better understand students, and the settings which they learn in [15]. Mining in educational environment is called Educational Data Mining, concern with developing new methods to discover knowledge from educational database[6].Lack of deep and enough knowledge in higher educational system may prevent system management to achieve quality objectives, data mining methodology can help this knowledge gaps in higher education system [11].

## VII. APPLICATION

In this study, data gathered from different degree colleges affiliated with Dr. R. M. L. Awadh University, Faizabad, India. These data are analyzed using classification method to predict the student's performance. In order to apply this technique following steps are performed in sequence:

1) **Data set**: The data set used in this study was obtained from different colleges on the sampling method of computer science department of course PGDCA of session 2009-10. Initially size of the data is 600.

2) **Database Software**: Microsoft SQL Server 2005 was used to store the data. The reason behind choosing MSSQL Server is: it was compatible and efficient to use with the database management system i.e. relational database.

3) **Application software**: Matlab environment is used as programming environment. The Matlab software suitable for the development of application with MSSQL server 2005.

4) **Data mining Process**: The data exploration and presentation process consisted of following steps:

   i) **Data preparation:** In this step the data that was maintained in different table was joined in a single table in 1 NF. After joining process all errors were removed.

   ii) **Data selection and transformation:** In this step only those fields were selected which were required for data mining. For example sex, language medium, stream of bachelor degree and division obtained. The data is transformed into the format of table 1. The head division shows the obtained division of the student in PGDCA final exam. Caste category shows the demographic distribution of the student defined by the GOI (Government of India). Language medium is the medium in which student passed his/her graduation program. Class is the stream which a student passed to get admission in PGDCA. This categorized as BA (NC), BA (CA), B Sc (Bio), B Sc (Math) and B Com. BA (NC) is for those BA students who did not take calculative subject in the BA program and BA (CA) is indicating those students who took calculative subject.

iii) **Implementation of mining model:** Given a training set the naïve Bayes algorithm first estimates the prior probability P (C$_j$) for each class by counting how often each class occurs in the training data. For ach attribute value x$_i$ can be counted to determine P (x$_i$). Similarly the probability P (x$_i$ | C$_j$) can be estimated by counting how often each value occurs in the class in the training data.

When classifying a target tuple, the conditional and prior probabilities generated

from the training set are used to make the prediction. Then estimate P (t$_i$ | C$_j$) by

$$P ( t_i \mid C_j) = \prod_{k=1}^{p} P( x_{ik} \mid C_j)$$

To calculate P (t$_i$) we can estimate the likelihood that t$_i$ is in each class. The probability that t$_i$ is in a class is the product of the conditional probabilities for each attribute value. The class with the highest probability is the one chosen for the tuple [10].

iv) **Results and discussion:** After Bayesian classification the table 1 of data produces the table 2 and figure 1 shows the pictorial representation of table2. Table 2 shows the highest probability of division of a particular class (medium, category and class). For example if a student is of medium, OBC category and BA with non-calculative subject then it can be predicted that s/he will score second division mark in the final exam.

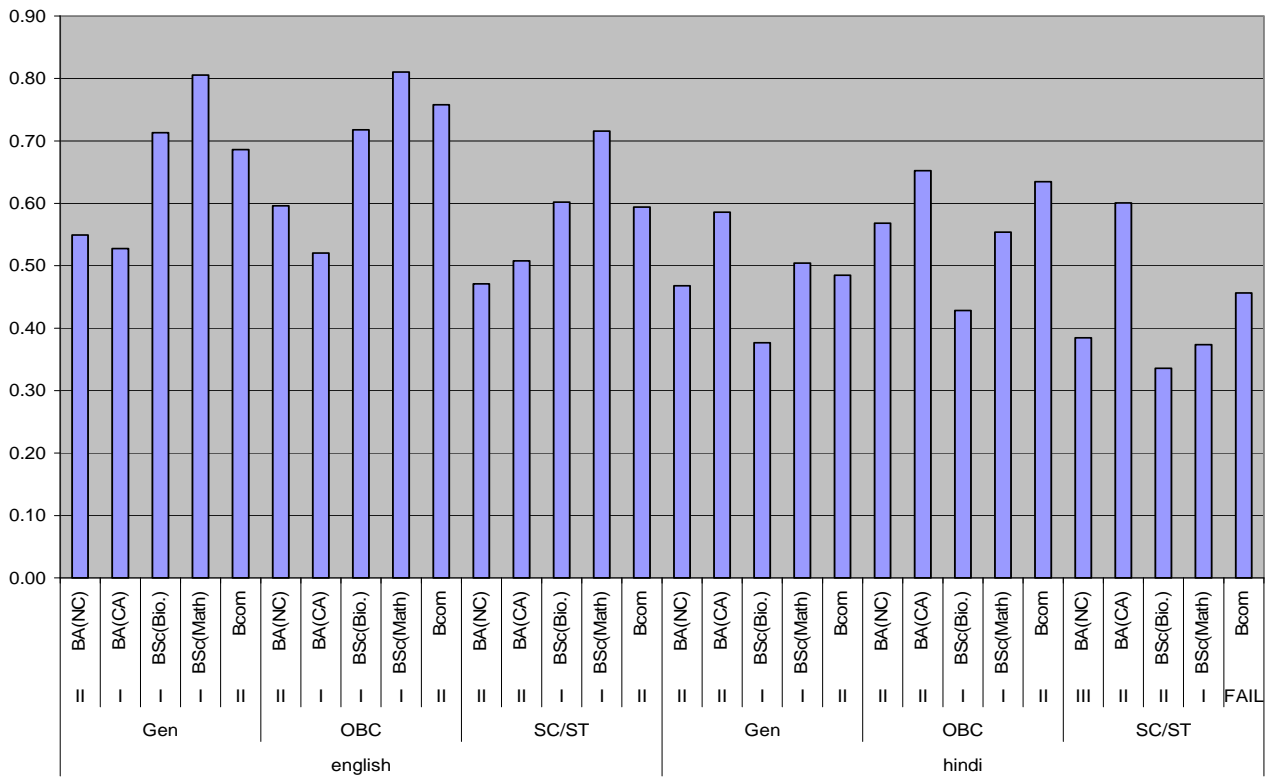

**Figure 1**

| Medium | Category | Division | Class | Probability |
|---|---|---|---|---|
| English | Gen | II | BA(NC) | 0.549218 |
| | | I | BA(CA) | 0.527331 |
| | | I | BSc(Bio.) | 0.712938 |
| | | I | BSc(Math) | 0.805456 |
| | | II | Bcom | 0.685979 |
| | OBC | II | BA(NC) | 0.596068 |
| | | I | BA(CA) | 0.520268 |
| | | I | BSc(Bio.) | 0.717771 |
| | | I | BSc(Math) | 0.810201 |
| | | II | Bcom | 0.758005 |
| | SC/ST | II | BA(NC) | 0.471241 |
| | | II | BA(CA) | 0.507795 |
| | | I | BSc(Bio.) | 0.601875 |
| | | I | BSc(Math) | 0.715802 |
| | | II | Bcom | 0.594066 |
| Hindi | Gen | II | BA(NC) | 0.467961 |
| | | II | BA(CA) | 0.585719 |
| | | I | BSc(Bio.) | 0.376518 |
| | | I | BSc(Math) | 0.504112 |
| | | II | Bcom | 0.484983 |
| | OBC | II | BA(NC) | 0.568261 |
| | | II | BA(CA) | 0.652269 |
| | | I | BSc(Bio.) | 0.428287 |
| | | I | BSc(Math) | 0.553675 |
| | | II | Bcom | 0.634676 |
| | SC/ST | III | BA(NC) | 0.384808 |
| | | II | BA(CA) | 0.600667 |
| | | II | BSc(Bio.) | 0.335702 |
| | | I | BSc(Math) | 0.373642 |
| | | FAIL | Bcom | 0.456478 |

**Table 2**

## VIII. CONCLUSION

In this paper, Bayesian classification method is used on student database to predict the students division on the basis of previous year database. This study will help to the students and the teachers to improve the division of the student. This study will also work to identify those students which needed special attention to reduce failing ration and taking appropriate action at right time.